\documentclass[preprint,12pt]{elsarticle}
\usepackage{graphicx}
\usepackage{amssymb}
\usepackage{amsmath}

\journal{Nuclear Physics A}

\begin{document}

\begin{frontmatter}

\title{$K^+$-nucleus potentials from $K^+$-nucleon amplitudes}

\author[a]{E.~Friedman\corref{cor1}}
\cortext[cor1]{Corresponding author: E. Friedman, elifried@cc.huji.ac.il}
\address[a]{Racah Institute of Physics, The Hebrew University, 91904
Jerusalem, Israel}

\begin{abstract}

Optical potentials for $K^+$-nucleus interactions are constructed from
$K^+$-nucleon amplitudes using recently developed algorithm based on
$K^+$-N kinematics in the nuclear medium. With the deep
penetration of $K^+$ mesons into the nucleus at momenta below 800~MeV/c
it is possible to test this approach with greater sensitivity than
hitherto done with $K^-$ and pions.
The energy-dependence of experimental
reaction and total cross sections on nuclei
is better reproduced with this approach
compared to fixed-energy amplitudes. The inclusion of Pauli correlations
in the medium also improves the agreement between calculation and experiment.
The absolute scale of the cross sections is reproduced very well for $^6$Li
but for C, Si and Ca calculated cross sections 
are (23$\pm4$)\% smaller than experiment, in agreement
with earlier analyses. Two phenomenological models that produce such
missing strength suggest that the imaginary part of the potential
needs about 40\% enhancement.

\end{abstract}


\end{frontmatter}

\section{Introduction and background}
\label{sec:intro}

The connection between hadron-nucleus interaction and the corresponding
hadron-nucleon interaction could provide
information on possible modifications of strong interaction properties in
the nuclear medium. The interaction of low energy pions with nuclei, and pionic
atoms in particular, is a good example where the dominance of the $P$-wave (3,3)
resonance in the pion-nucleon interaction underlies the Kisslinger \cite{Kis55} 
and the Ericson-Ericson~\cite{EEr66} models for the pion-nucleus interaction. 
The depth of
penetration of a meson into the nucleus is a key element in this connection
and indeed below 80~MeV empirical pion-nucleus potentials 
display the characteristics of the $P$-wave model~\cite{Fri83}. 
As the pion energy increases towards the resonance with the associated 
increased absorption,
it is found that good description of the pion-nucleus interaction is possible 
also without
an explicit $P$-wave model \cite{Sat92,JSa96}. 

The present work deals with $K^+$ meson-nucleus interaction at 
low energies where
the total $K^+$-nucleon cross sections are small and the kaon penetrates
well into the nucleus. This property of the $K^+$ is well known \cite{BGK68}
 and some experiments in the 1990s were
 motivated by it \cite{Mar90,Kra92,Saw93,Wei94}. 
The present study was inspired by recent works
\cite{CFG11,CFG11a,FGa12,GMa12,FGa13,FGM13,CFG14,FGa14,FGL15}
where algorithms based on {\it in-medium kinematics} have been applied in
calculations of optical potentials from the corresponding hadron-nucleon
scattering amplitudes.

Applying this approach to $K^+$ mesons,
the kaon-nucleon scattering amplitude is presented in terms of the Mandelstam
variable $s$ in the nuclear medium

\begin{equation}
\label{eq:s}
s = (E_K + E_N)^2 -(\vec p _K + \vec p_N)^2,
\end{equation}
\noindent
where \noindent
$E_K = m_K + E_{\rm lab},~~E_N = m_N - B_N$. 
$E_{\rm lab}$ is the laboratory
kinetic energy of the kaon and $B_N$ is an average binding energy of a nucleon.
In the nuclear medium
$\vec p _K + \vec p_N \ne $ 0 where $\vec p _K $ is determined by the beam
energy and the optical potential, 
$\vec p_N$ is determined by the nuclear environment.
Averaging over angles,
$(\vec p _K + \vec p_N)^2 \rightarrow (p _K)^2 + (p_N)^2.$
Substituting locally

\begin{equation}
\label{eq:pKsq}
 p_K^2/2 m_K \rightarrow  E_{\rm lab} - {\rm Re}~V^K_{\rm opt}-V_c, 
\end{equation}
with $V^K_{\rm opt}$ the kaon-nucleus optical potential and $V_c$ the Coulomb
potential. For the nucleon the Fermi gas model yields

\begin{equation}
\label{eq:pNsq}
p_N^2/2 m_N \rightarrow T_N (\rho /\bar \rho)^{2/3}  
\end{equation}
with $T_N$=23 MeV the average kinetic energy, $\rho $ and $\bar \rho$ the
local and average nuclear densities, respectively.

The simplest
`$t \rho$' form  of the optical potential is \cite{FGa07}

\begin{equation}
\label{eq:potl}
2\epsilon_{\rm red}^{(A)} V_{{\rm opt}}^K(r) = -4\pi F_A b_0(\sqrt s)\rho(r)
\end{equation}
where $\epsilon_{\rm red}^{(A)}$ is the c.m. reduced energy
\begin{equation}
\label{eq:ered}
(\epsilon_{\rm red}^{(A)})^{-1}=E_p^{-1} + E_A^{-1}
\end{equation}
in terms of the total energies $E_p$ for the projectile and $E_A$ for
the target and
\begin{equation}
\label{eq:FA}
F_A=\frac{M_A \sqrt s}{m_N (E_A+E_p)}
\end{equation}
is a kinematical factor resulting from the transformation of amplitudes
between the projectile-nucleon and the projectile-nucleus systems. 
 $b_0$ is the isospin-averaged free
$K^+$-N forward scattering amplitude.

Defining $\delta \sqrt s =\sqrt s -E_{\rm th}$ with $E_{\rm th}=m_{K}+m_N$,
then to first order in $B/E_{\rm th}$ and $(p/E_{\rm th})^2$ 
one gets \cite{FGa13}
\begin{equation}
\label{eq:sqrts}
\delta \sqrt s -\xi_{N}E_{\rm lab}= -B_N\rho/{\bar \rho}
-\xi_N T_N(\rho/\bar \rho)^{2/3}
 +\xi_{K}[{\rm Re}~V_{\rm opt}+V_c (\rho/\rho_0)^{1/3}],
\end{equation}
with $\xi_N=m_N/(m_N+m_{K}),~\xi_{K} = m_{K}/(m_N+m_{K})$.
This value of $\sqrt s$ serves as the argument of the {\it in 
medium} kaon-nucleon
amplitude in constructing the kaon-nucleus optical potential.

\begin{figure}[htb]
\begin{center}
\includegraphics[height=70mm,width=0.75\textwidth]{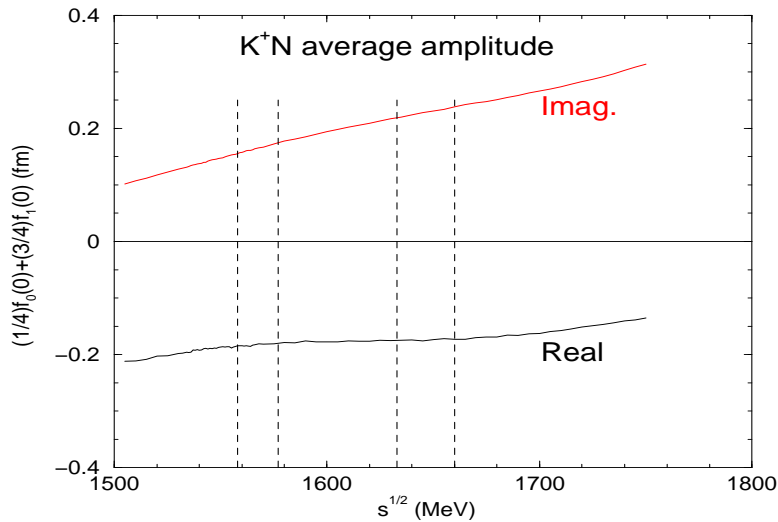}
\caption{Isospin-averaged $K^+$-nucleon amplitudes. Vertical dashed lines
indicate the four experimental energies discussed in the present work.}
\label{fig:free}
\end{center}
\end{figure}

Figure~\ref{fig:free} shows the isospin-averaged $K^+$-nucleon amplitude
as function of the c.m. energy $\sqrt s$, taken from the
SAID software package \cite{SAID} including $S,P$ and $D$ partial waves. 
Vertical dashed lines indicate the energies
corresponding to the four experimental energies considered in the present work.
It is seen that the variation of the
imaginary part of the amplitude is significant and could result in
observable effects within the above algorithm.

With the relatively small $K^+$-N total cross sections below 800~MeV/c
kaons penetrate deeply into nuclei and the above algorithm can be tested  
with greater sensitivity than before. In analyses of pionic 
atoms since true
absorption is impossible on a single nucleon, a two-nucleon
phenomenological term is always present in the potential that fits 
experimental results. 
Similarly phenomenological two-nucleon terms are found in analyses
of elastic scattering of $\pi ^\pm$ by nuclei at low energies.
For $K^-$ atoms multi-nucleon terms
are found when comparing calculations with  experiment. For $\eta$
mesons there are no experimental results yet for $\eta$-nucleus
interaction to compare with. For anti-protons there is a vast collection
of experimental results but the very strong absorption precludes a real
test of the above algorithm in the nuclear medium. Consequently
the  $K^+$-nucleus below 800~MeV/c is a preferred probe because there are
experimental results to compare with and there is no evidence that
there are significant  multi-nucleon contributions. In the present work
we use mostly reaction and total cross sections 
on nuclei as these have been 
measured to good accuracy. Few experimental angular distributions for
elastic scattering of $K^+$ by C are discussed below.

In section \ref{sec:data} we briefly review past work on the topic 
of $K^+$-nucleus integral cross sections and in section \ref{sec:results}
we apply the above approach to compare between experimental results
and calculations. Section \ref{sec:summ} is summary and conclusions.

\section{$K^+$-nucleus integral cross sections}
\label{sec:data}

The pioneering experiment of Bugg et al. \cite{BGK68} showed that  
 above 800~MeV/c total cross sections for $K^+$ on carbon were smaller than
six times the corresponding cross sections on deuteron, as expected.
In contrast, below that momentum the reverse was true, exhibiting enhancement
of the nuclear total cross section relative to the nucleon cross section. This 
statement obviously depends on the assumption that nuclear medium effects on the
deuteron are very small. Later Siegel, Kaufmann and Gibbs \cite{SKG85} showed
that calculations of the ratio $K^+$-carbon to $K^+$-
deuteron total cross sections which included traditional
medium corrections and their uncertainties, failed to reproduce the above
enhancement, thus suggesting more exotic mechanisms. Several publications that
followed considered density-dependent effective masses of the vector mesons
\cite{BDS88}, meson exchange current \cite{JKol92}, pion cloud contribution
\cite{GNO95} and mesons exchanged between the $K^+$ and the target nucleons
\cite{CLa96}.

The experimental situation with $K^+$-nucleus interactions changed when in the 
early 1990s total cross sections for D, $^6$Li, $^{12}$C, Si and Ca 
were measured
at four momenta below 800 MeV/c \cite{Mar90,Kra92,Saw93,Wei94}. The method was 
to measure transmissions through a target using detectors subtending different
solid angles and extrapolating to
zero solid angle after applying  model-dependent corrections. The same set
of transmission measurements was later reanalysed  
\cite{FGW97} to determine not only total
cross sections but also total reaction cross sections that are less dependent on
applied corrections. That increased the number of measured integral cross sections
from 16 to 32, in addition to those on D. Self-consistent analyses of this set
of data \cite{FGM97a,FGM97b} confirmed that the cross sections on C, Si and Ca
were 15-20\% larger than expected compared to D and to the very low-density
nucleus $^6$Li. Phenomenological fits to the data required mainly increased
imaginary part of the effective $K^+$-nucleon amplitudes in the medium.
Analysis of the same data by Peterson \cite{Pet99} in terms
of an average $S$-wave phase-shift showed good agreement with experiment when
increasing this phase-shift in the medium relative to free space.
A possible link between the additional reactive content and $K^+$ interactions
with two nucleons was explored in \cite{GFr05,GFr06}.

\section{Results}
\label{sec:results}

\begin{table}[htb]
\caption{Comparisons between total cross sections (in mb)
of $K^+$ on $^6$Li with three times the cross sections on deuteron. Only
statistical errors are quoted.}
\label{tab:DLi}
\begin{center}
\begin{tabular}{ccccc}
$p_{{\rm lab}}$ (MeV/c) & 488 & 531 & 656 & 714 \\ \hline
3$\times$D &76$\pm$1.8 & 81.5$\pm$ 1.0&84.5$\pm$ 0.7&86.0$\pm$ 0.6 \\
$^6$Li & 77.5$\pm$1.1 & 80.7$\pm$ 0.7&86.4$\pm$ 0.7&88.5$\pm$ 0.6 \\ \hline
\end{tabular}
\end{center}
\end{table}

 The experimental results used in the present analysis are reaction
and total cross sections for $K^+$ on $^6$Li, C, Si and Ca
from Ref.\cite{FGM97b}. Total cross sections
on D are taken from Ref.\cite{FGW97}. Table \ref{tab:DLi} compares 
total cross sections of $K^+$ on $^6$Li with three times the cross sections
on deuteron. It is seen that within 2\% there are no indications for
medium effects in the results for $^6$Li, where the average density is
about half of the average density for other nuclei. Consequently in what
follows we use $^6$Li as a reference.

\begin{figure}
\begin{center}
\includegraphics[height=70mm,width=0.75\textwidth]{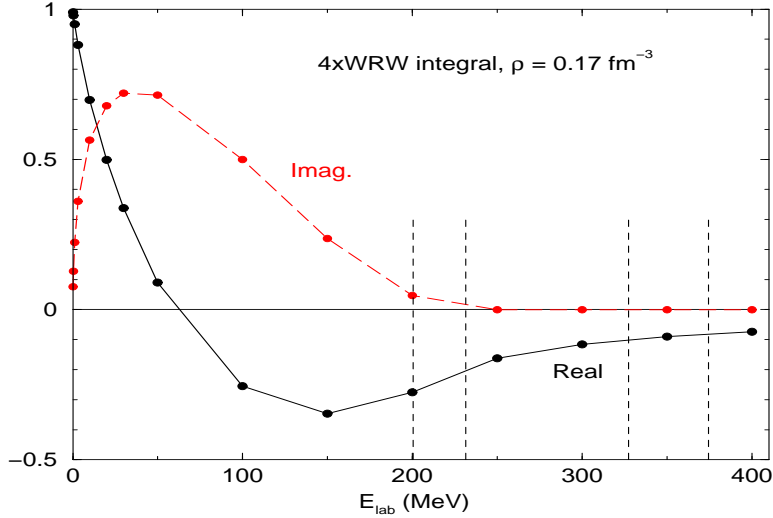}
\caption{Four times the integral of Eq.(\ref{eq:WRW}) as function of
energy for a full nuclear density.
Vertical dashed lines indicate the experimental energies.}
\label{fig:WRWint}
\end{center}
\end{figure}

\begin{figure}
\begin{center}
\includegraphics[height=70mm,width=0.75\textwidth]{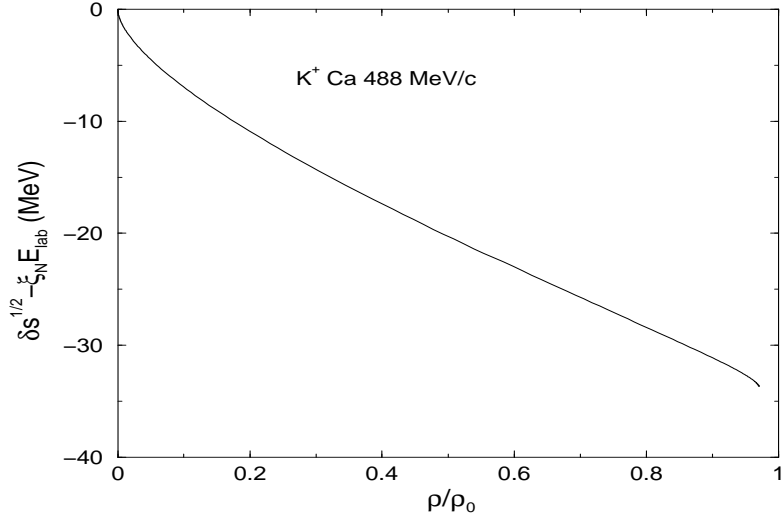}
\caption{Example of density to energy transformation implied by
Eqs.(\ref{eq:potl}) and (\ref{eq:sqrts}).}
\label{fig:delta21}
\end{center}
\end{figure}

\subsection{Testing in-medium kinematics}
\label{sec:testing}
Before applying Eq.(\ref{eq:sqrts}) to calculate $K^+$-nucleus cross sections
it is necessary to correct
 the free $K^+$-nucleon amplitudes by considering Pauli 
correlations. The corrections of Waas, Rho and Weise (WRW) \cite{WRW97}
replace $F_A b_0 \rho$ of Eq.(\ref{eq:potl}) by
\begin{equation}
\label{eq:WRW0}
\sum_{I}\frac{2I+1}{4}\;
\frac{{F_A f}_{I}}{1+\frac{1}{4}\xi{F_A f}_{I}\rho(r)}\:\rho(r)
\end{equation}
where $f_I$ are the isospin $I$ free $K^+$-nucleon 
forward scattering amplitudes.
Therefore the  medium corrections are determined by the quantity
\begin{equation}
\label{eq:WRW}
\xi=\frac{9\pi}{p_F^2}\left( 4\int_0^{\infty}{\frac{dt}{t}\,\exp({\rm i}qt)\,
j_1^2(t)} \right) \; , \;\;\;\;\;\;\; q=k/p_F \;.
\end{equation}
where $k=(\omega_K^2-m_K^2)^{1/2}$ with $\omega_K$ the c.m. energy
and the Fermi momentum $p_F$  
given by $p_F=(3{\pi}^2\rho/2)^{1/3}$. For kaonic atoms $k\approx 0$ and the expression
in brackets is 1, but for scattering ($k\ne 0$)  
the integral (and the corrections)
go down rapidly with increasing energy. 
Figure~\ref{fig:WRWint}
shows as an example, four times the above integral as function 
of energy for a Fermi momentum corresponding to central nuclear density.

\begin{figure}
\begin{center}
\includegraphics[height=85mm,width=0.75\textwidth]{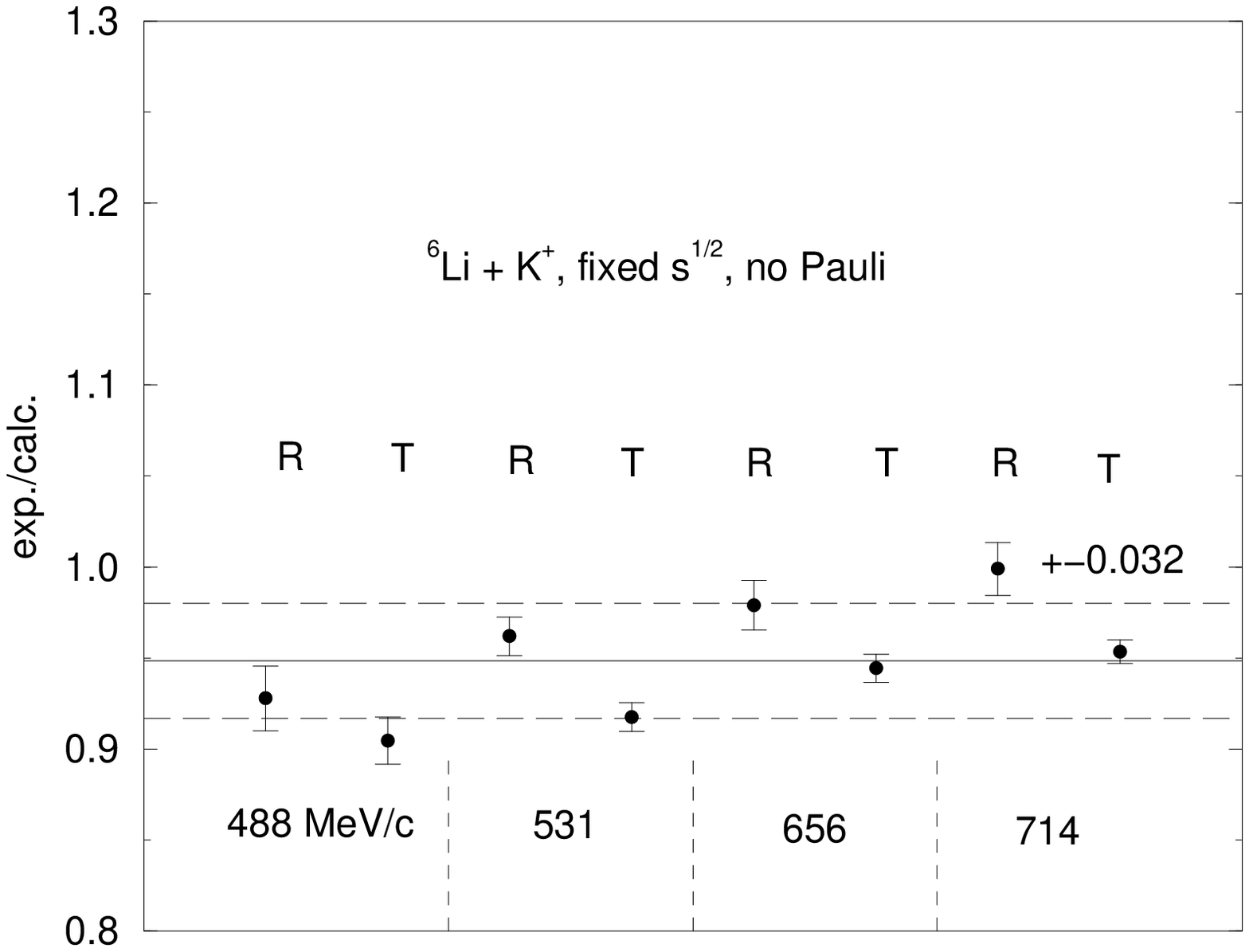}
\includegraphics[height=85mm,width=0.75\textwidth]{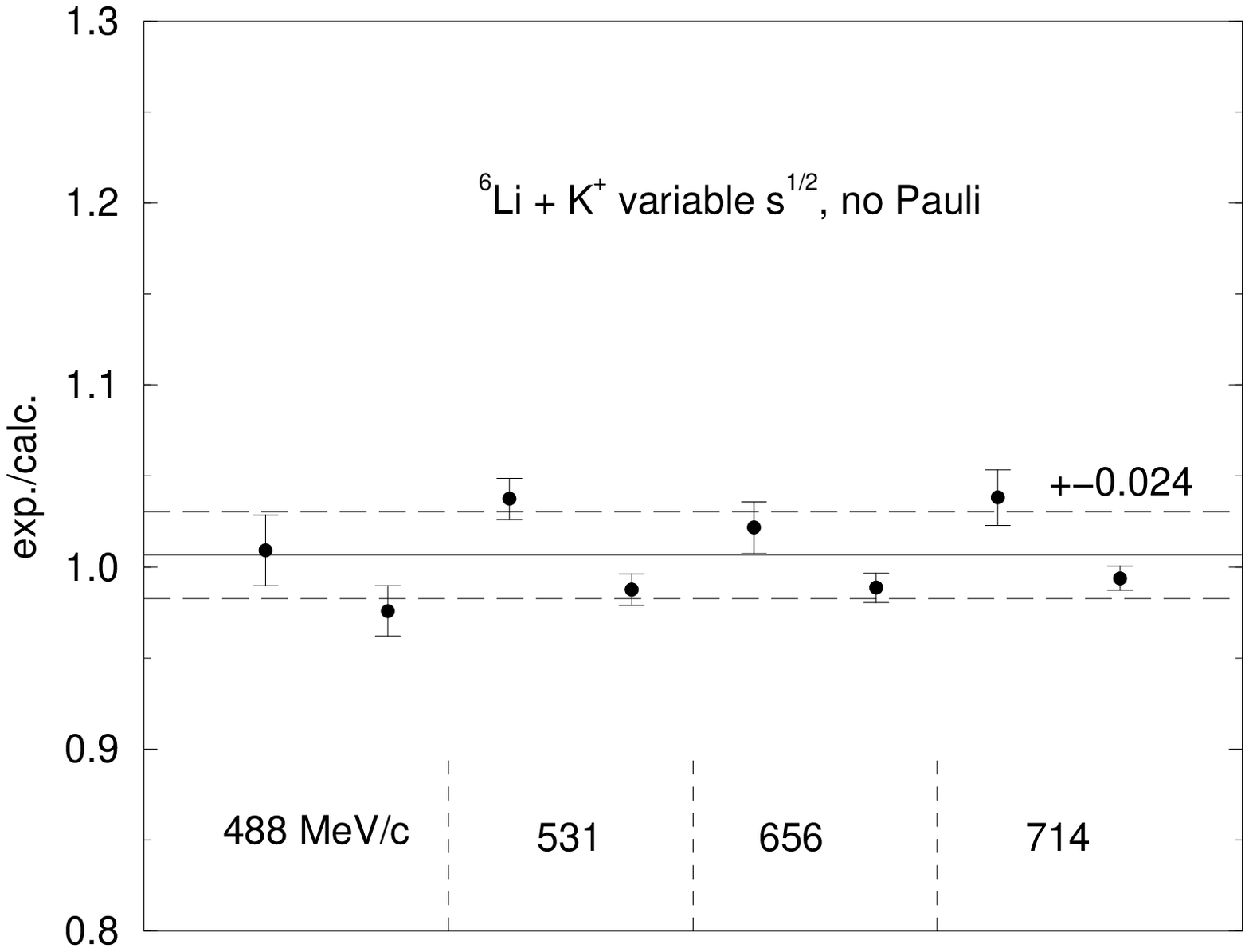}
\caption{Experimental to calculated ratios of reaction and total cross 
sections on $^6$Li for fixed energies (top)
and for variable energies Eq.(\ref{eq:sqrts}) (bottom). R and T indicate
reaction and total cross sections, respectively.}
\label{fig:Lifixed}
\end{center}
\end{figure}

Equations (\ref{eq:potl}) and (\ref{eq:sqrts}) define a density
to energy transformation through the scattering amplitudes.
As the real part of the potential determines the energy and, in turn,
the energy and density determine the amplitudes, a self-consistent solution
is required. Good convergence is usually achieved after 4-5 iterations
and an example for this transformation is shown in Fig.~\ref{fig:delta21}
for 488~MeV/c $K^+$ interacting with Ca. 
An energy range of 30 MeV that corresponds
to penetration of the kaon into regions of up to full nuclear
density could be significant, as
can be seen from Fig.~\ref{fig:free}

\begin{figure}
\begin{center}
\includegraphics[height=85mm,width=0.75\textwidth]{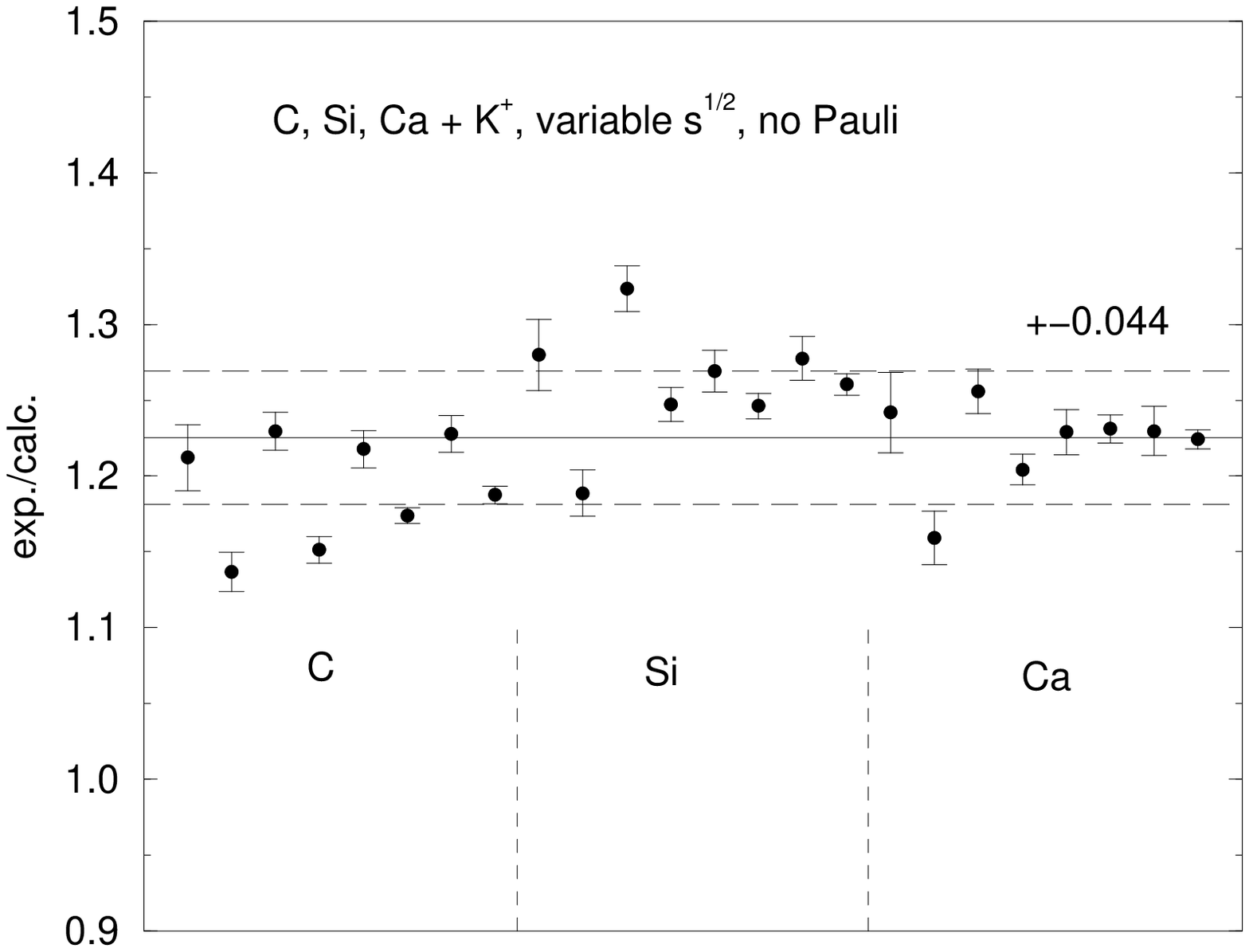}
\includegraphics[height=85mm,width=0.75\textwidth]{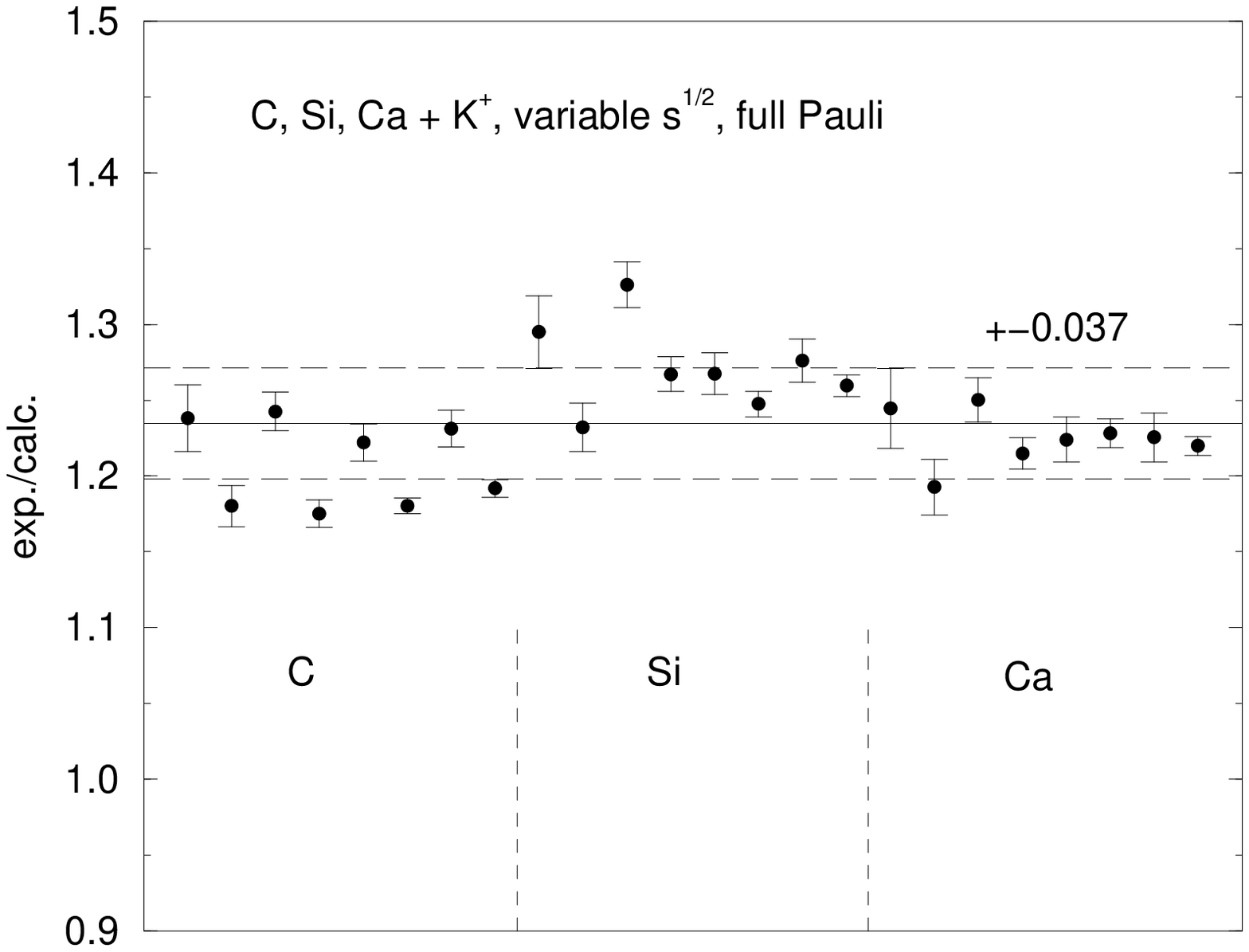}
\caption{Experimental to calculated ratios of reaction and total cross
sections on C, Si and Ca for variable energies
Eq.(\ref{eq:sqrts}) (top), and with Pauli correlations also included (bottom).}
\label{fig:Nucvar}
\end{center}
\end{figure}

As a first step in comparisons between calculation and experiment we show
in the top part of Fig.~\ref{fig:Lifixed}  reaction and total 
cross sections on $^6$Li.
The calculations use amplitudes at {\it fixed} energies related to
the beam energies without respecting Eq.(\ref{eq:sqrts})
and without the Pauli correlation correction (WRW). Ratios of experimental
to calculated cross sections are shown for the four energies, alternating
between reaction (R) and total (T) cross sections. 
It is gratifying that to within 5\% the
calculations reproduce the absolute scale of the experimental results.
However, upon closer examination it is clear that there is a systematic
difference between calculated and experimental dependence on energy of the
cross sections. 
The bottom part of  Fig.~\ref{fig:Lifixed}  
shows similar results for calculations
based on the full variable energy of Eq.(\ref{eq:sqrts}) and it is evident
that the energy dependence is reproduced very well by calculations. 
The value of $T_N$ (Eq.(\ref{eq:sqrts})) was adjusted to the low density
of $^6$Li. Adding also
the Pauli correlations made very small difference.
It is therefore seen that (i) From the energy-dependence of the ratios 
the prescription of Eq.(\ref{eq:sqrts})
is supported by experiment and (ii) Effects due to Pauli correlations
are very small for the low-density $^6$Li nucleus.

Turning to the other target nuclei, namely C, Si and Ca, calculations
with fixed energies (not shown) display systematic deficiencies regarding
the variation with energy of the experimental to calculated 
ratios, similarly 
to the results for $^6$Li. Figure~\ref{fig:Nucvar} (top) shows
the ratios obtained using the $\sqrt s$ algorithm Eq.(\ref{eq:sqrts}) but
without the Pauli correlations corrections of WRW. 
The energy dependence of the ratios for the reaction cross sections is
seen to be well reproduced by the calculations
but there is a residual disagreement in the energy dependence for
the total cross sections, notably for C. 
The bottom part of Fig.~\ref{fig:Nucvar}
 shows these ratios where the WRW in-medium corrections
have also been included, and the improved agreement is evident. The larger
scatter of the ratios for Si has been observed also in earlier analyses
\cite{FGM97b}.

\subsection{In-medium enhancement}
\label{sec:enhancement}

Agreement to 2-3\% between calculation and experiment for the reaction
and total cross sections on the very low density $^6$Li, 
as obtained above, had not
been observed in earlier analyses. Consequently we
 deal here directly with the cross sections for the `normal' targets of
C, Si and Ca, without the use of `super-ratios', namely, without
normalizing experimental to calculated ratios for a given target to the
corresponding ratios for $^6$Li. Moreover, with the present model for 
variations of the in-medium energies we try to fit calculations to
experiment for the three targets at the four energies put together,
a total of 24 data points.

The first phenomenological approach is to simply multiply separately
the real and the imaginary parts of the amplitudes by a respective
scaling factor, the same for all energies and four targets.
Obviously no fit to the data is possible with $^6$Li included 
because for $^6$Li no
scaling factors are expected whereas this is not the case
for C, Si and Ca. Removing $^6$Li from the data, fits of two parameters
to the three heavier targets yields for
24 data points $\chi ^2$=23.8 when using assigned errors
of $\pm$3.7\% as implied by the scatter of the ratios of
calculated to experimental cross
sections (see Fig. \ref{fig:Nucvar}). The scaling factors are then
$F_{\rm R}$=0.87$\pm$0.10 for the real part 
and $F_{\rm I}$=1.41$\pm$0.02 for the 
imaginary part of the potential.
A separate fit to the eight $^6$Li cross sections 
leads to $\chi ^2$=8.7 with
scaling factors of 0.71$\pm$0.10 and 1.02$\pm$0.02 for the real 
and imaginary parts, respectively. 
Reducing the assigned errors all the way down to the statistical errors
does not change the resulting scaling factors. Consequently
we conclude that the real part of the amplitudes
needs rescaling by 0.9$\pm$0.1 whereas the imaginary part needs
rescaling by 1.40$\pm$0.02 in order to fit the reaction
and total cross sections on C, Si and Ca.

Alternatives to rescaling of the input amplitudes have been tried
too by adding a phenomenological term to the potential in order
to fit the C, Si and Ca cross sections. Fits were possible by
adding a term with higher powers of density than linear, 
to simulate two-nucleon processes.
Another version was to add a linear term as a
$P$-wave potential of the Kisslinger type \cite{Kis55}. Both options
could close the 23\% gap between the calculated cross sections based on the
input amplitudes and experiment. However, comparisons with the few available
results on the elastic scattering of $K^+$ by C did not support such
 approaches. 

\begin{figure}[htb]
\begin{center}
\includegraphics[height=85mm,width=0.75\textwidth]{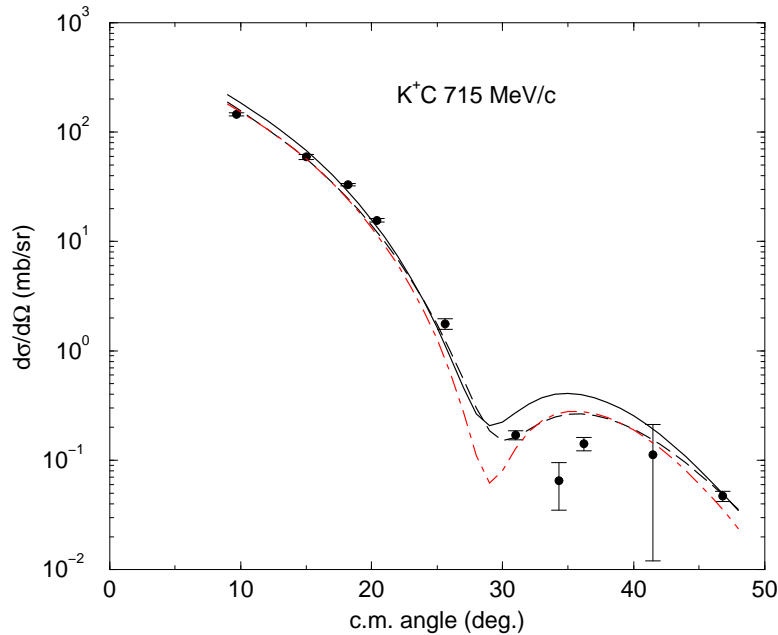}
\caption{Calculated and experimental differential cross sections for elastic
scattering of 715 MeV/c $K^+$ by C.
Experimental results are from Ref. \cite{CSP97}. Dashed curve for
amplitudes without
rescaling, solid curve for rescaling by constants,
dot-dashed curve for added phenomenological $p$-wave potential, see text.}
\label{fig:angdist}
\end{center}
\end{figure}

Out of the few experiments below 800 MeV/c we chose elastic scattering
of $K^+$ by C at 715~MeV/c \cite{CSP97}. In Fig. \ref{fig:angdist}
the dashed curve shows the predictions made using the present model
with in-medium kinematics and WRW Pauli correlations effects. The
solid curve shows the results when the above rescaling factors 
$F_{\rm R}$ and $F_{\rm I}$
are applied to the real and imaginary parts, respectively, to fit the
(enhanced) reaction and total cross sections. Both are in fair
agreement with experiment, showing very little sensitivity of the 
angular distribution to modifications required to fit the reaction 
and total cross sections. The option of adding a phenomenological
non-linear term or a linear $P$-wave term to the potential 
(not shown), produced
angular distributions that disagree sharply with experiment beyond the
first minimum.  Another attempt was to use for the in-medium 
forward scattering amplitudes
only the $S$-wave amplitudes from SAID \cite{SAID}, thus neglecting
the contributions from the $P$- and $D$-wave terms, and to add 
an explicit phenomenological $P$-wave potential \cite{Kis55}. 
That too made it possible to
fit the reaction and total cross section with only two parameters.
In this case the balance between repulsive and attractive real parts 
of the potential changed and
the predicted angular distribution, dot-dashed curve
in Fig. \ref{fig:angdist}, is acceptable.
Very similar results were obtained with $K^+$ elastically
scattered by C at 635 MeV/c \cite{CSP97}.

We have also checked if it was possible to include $^6$Li in the data
and to fit all 32 cross sections with
additional terms in the potential.
 This approach failed before
\cite{FGM97a,FGM97b} and it has likewise failed in the present work.
However, it was shown in Ref. \cite{FGM97a,FGM97b} that a
possible way to handle $^6$Li and the
other targets together was by using as a parameter an {\it average} density
of the nucleus that turns out to cut-off $^6$Li from the rescaling
of the imaginary potential. In this
model the real part of the potential is rescaled by a constant
whereas the imaginary part is multiplied by the expression
\begin{equation}
\label{eq:theta}
F_{\rm I}=[1+{\rm Im}B_0(\bar \rho -\rho _c) \Theta(\bar \rho -\rho _c)]
\end{equation}
with $\bar \rho$ the average
density of the target nucleus and $\rho _c$ a cut-off density.
With $\rho _c$ turning out to be
0.088$\pm$0.005 fm$^{-3}$, larger than the average density
of the $^6$Li nucleus, it was possible to fit all 32 data points
with only three parameters. The predicted angular distribution for
elastic scattering of 715~MeV/c K$^+$ by C 
when using this prescription is indistinguishable
from the solid curve in Fig. \ref{fig:angdist}.

\section{Discussion and summary}
\label{sec:summ}

Recent in-medium algorithm for calculating optical potentials from the
respective meson-nucleon 
forward scattering amplitudes, used so far for kaonic atoms, pionic
atoms, low energy pion scattering and $\eta$-nucleus interactions, was 
used in the present work to calculate integral cross sections for
 low energy $K^+$ mesons. 
This work was motivated by 
 better penetrability of kaons into nuclei below 800~MeV/c
in comparison with other particles thus providing
 more sensitive tests of this approach. 
In particular, penetration beyond the nuclear surface into where the
density is close to the full nuclear density means shifts of up to
30~MeV in the argument of the in-medium amplitude which causes
observable effects.
It was possible to test
separately the effects of the energy-dependence implied by the model
and the importance of applying corrections due to Pauli correlations.
Both were found to improve the agreement between experiment and calculations
regarding the energy-dependence of total and reaction cross sections of
$K^+$ on C, Si and Ca, particularly at the lower energy end of the range. 

For the low-density $^6$Li nucleus, calculation and experiment 
agree to better than 3\% uniformly over the energy range studied
when the variable energy of the model is applied. Effects due to
Pauli correlations are negligibly small. 
Compared with earlier analysis \cite{FGW97,FGM97a,FGM97b} the full 
agreement achieved for $^6$Li is partly due to the inclusion in the present
work of $D$-wave contributions to the forward scattering amplitude. 
The $D_{03}$ partial wave contributes attraction that
is close to 30\% of the repulsion due to the $S_{11}$ partial wave,
particularly at the lower energies.

The $^6$Li 
provides a solid basis for direct comparisons with the other three
nuclear targets without a need to use so-called `super-ratios', namely,
normalizing experimental to calculated ratios for a target to the
corresponding ratios for $^6$Li. The experimental cross sections  
for C, Si and Ca are found to be (23$\pm$4)\%  larger than calculation,
in general agreement with previous observations.
This enhancement could be reproduced phenomenologically
by rescaling the imaginary part of the potential by 40$\pm$2\% but
quantitative understanding of this effect could not be achieved.
It suggests that  
in-medium enhancements of $K^+$-nucleon interaction are 
due to more exotic
mechanisms than traditional corrections. Comparisons with predictions
of such mechanisms are obviously called for.

\section*{Acknowledgements}

Fruitfull discussions with A.~Gal are gratefully acknowledged.


\begin{thebibliography}{99}

\bibitem{Kis55} L.S.~Kisslinger, Phys. Rev. 98 (1955) 761

\bibitem{EEr66} M.~Ericson, T.E.O.~Ericson, Ann. Phys. 36 (1966) 323.

\bibitem{Fri83} E.~Friedman, Phys. Rev. C 28 (1983) 1264.

\bibitem{Sat92} G.R.~Satchler, Nucl. Phys. A 540 (1992) 533.

\bibitem{JSa96} M.B.~Johnson, G.R.~Satchler, Ann. Phys. 248 (1996) 134.

\bibitem{BGK68} D.V.~Bugg et al., Phys. Rev. 168 (1968) 1466.

\bibitem{Mar90} Y.~Mardor et al., Phys. Rev. Lett. 65 (1990) 2110.

\bibitem{Kra92} R.A.~Krauss et al., Phys. Rev. C 46 (1992) 655.

\bibitem{Saw93} R.~Sawafta et al., Phys. Lett. B 307 (1993) 293.

\bibitem{Wei94} R.~Weiss et al., Phys. Rev. C 49 (1994) 2569

\bibitem{CFG11} A.~Ciepl\'{y}, E.~Friedman, A.~Gal, D.~Gazda, J.~Mare\v{s},
Phys. Lett. B 702 (2011) 402.

\bibitem{CFG11a} A.~Ciepl\'{y}, E.~Friedman, A.~Gal, D.~Gazda, J.~Mare\v{s},
Phys. Rev. C 84 (2011) 045206.

\bibitem{FGa12} E.~Friedman, A.~Gal, Nucl. Phys. A 881 (2012) 150.

\bibitem{GMa12} D.~Gazda, J.~Mare\v{s}, Nucl. Phys. A 881 (2012) 159.

\bibitem{FGa13} E.~Friedman, A.~Gal, Nucl. Phys. A 899 (2013) 60.

\bibitem{FGM13} E.~Friedman, A.~Gal, J.~Mare\v{s}, Phys. Lett. B 725 (2013)
334.

\bibitem{CFG14} A.~Ciepl\'{y}, E.~Friedman, A.~Gal, J.~Mare\v{s},
Nucl. Phys. A 925 (2014) 126.

\bibitem{FGa14} E. Friedman, A. Gal, Nucl. Phys. A 928 (2014) 128.

\bibitem{FGL15} E.~Friedman, A.~Gal, B.~Loiseau, S.~Wycech, 
Nucl. Phys. A 943 (2015) 101.

\bibitem{FGa07} E.~Friedman, A.~Gal, Phys. Rep. 452 (2007) 89.

\bibitem{SAID} http://gwdac.phys.gwu.edu

\bibitem{SKG85} P.B.~Siegel, W.B.~Kaufmann, W.R.~Gibbs, Phys. Rev. C 31 (1985) 2184.

\bibitem{BDS88} G.E.~Brown, C.B.~Dover, P.B. Siegel, W.~Weise, Phys. Rev.
                Lett. 60 (1988) 2723.

\bibitem{JKol92} M.F.~Jiang,  D.S.~Koltun, Phys. Rev. C 46 (1992) 2462.

\bibitem{GNO95} C.~Garcia-Recio, J.~Nieves, E.~Oset, Phys. Rev. C 51 (1995) 237.

\bibitem{CLa96} J.C.~Caillon, J.~Labarsouque, Phys. Rev. C 53 (1996) 1993.

\bibitem{FGW97} E.~Friedman et al., Phys. Rev. C 55 (1997) 1304.

\bibitem{FGM97a} E.~Friedman, A.~Gal, J.~Mare\v{s}, Phys. Lett. B. 396 (1997) 21.

\bibitem{FGM97b} E.~Friedman, A.~Gal, J.~Mare\v{s}, Nucl. Phys. A 625 (1997) 272.

\bibitem{Pet99} R.J.~Peterson, Phys. Rev. C 60 (1999) 022201.

\bibitem{GFr05} A.~Gal, E.~Friedman, Phys. Rev. Lett. 94 (2005) 072301.

\bibitem{GFr06} A.~Gal, E.~Friedman, Phys. Rev. C 73 (2006) 015208.

\bibitem{WRW97} T.~Waas, M.~Rho, W.~Weise, Nucl. Phys. A 617 (1997) 449.

\bibitem{CSP97} R.E.~Chrien, R.~Sawafta, R.J.~Peterson, R.A.~Michael,
E.V.~Hungerford, Nucl. Phys. A 625 (1997) 251.

\end{thebibliography}
\end{document}